\documentclass[12pt,preprint]{aastex}
\usepackage{color}

\newcommand{\bdv}[1]{\mbox{\boldmath$#1$}}

\def\au{{\rm AU}}

\def\masyr{{\rm mas}\,{\rm yr}^{-1}}
\def\kpc{{\rm kpc}}
\def\mas{{\rm mas}}

\def\muas{\mu{\rm as}}

\def\rel{{\rm rel}}

\def\eff{{\rm eff}}

\def\e{{\rm E}}
\def\bpi{{\bdv\pi}}
\def\bmu{{\bdv\mu}}

\begin{document}
\title{OGLE-2017-BLG-0173Lb: Low Mass-Ratio Planet in a ``Hollywood'' Microlensing Event}

\author{\textsc{
K.-H. Hwang$^{1}$,  
A. Udalski$^{2}$,
Y.~Shvartzvald$^{3,^{\dag}}$, 
Y.-H. Ryu$^{1}$, 
\and 
M. D. Albrow$^{4}$, 
S.-J. Chung$^{1,5}$, 
A. Gould$^{1,6,7}$, 
C. Han$^{8}$, 
Y. K. Jung$^{9}$ ,
I.-G. Shin$^{9}$, 
J. C. Yee$^{9}$, 
W. Zhu$^{6}$, 
S.-M. Cha$^{1,10}$, 
D.-J. Kim$^{1}$, 
H.-W. Kim$^{1}$, 
S.-L. Kim$^{1,5}$, 
C.-U. Lee$^{1,5}$,
D.-J. Lee$^{1}$,
Y. Lee$^{1,10}$, 
B.-G. Park$^{1,5}$,
R. W. Pogge$^{6}$ \\
(KMTNet Collaboration)\\
J. Skowron$^{2}$, 
P. Mr\'{o}z$^{2}$, 
R. Poleski$^{2,6}$,
S. Koz{\l}owski$^{2}$, 
I. Soszy\'{n}ski$^{2}$, 
P. Pietrukowicz$^{2}$, 
M. K. Szyma\'{n}ski$^{2}$, 
K. Ulaczyk$^{2}$,
M. Pawlak$^{2}$\\
(OGLE Collaboration)\\
G.~Bryden$^{3}$, 
C. Beichman$^{11}$, 
S. Calchi Novati$^{12}$,
B. S. Gaudi$^{6}$, 
C. B. Henderson$^{11}$,
S.~Jacklin$^{13}$,
M. T. Penny$^{6,^{\dag\dag}}$\\
(UKIRT Microlensing Team)\\}}

\affil{$^{1}$Korea Astronomy and Space Science Institute, Daejon
34055, Korea}

\affil{$^{2}$Warsaw University Observatory, Al. Ujazdowskie 4,
00-478 Warszawa, Poland}

\affil{$^{3}$Jet Propulsion Laboratory, California Institute of
Technology, 4800 Oak Grove Drive, Pasadena, CA 91109, USA}

\affil{$^{4}$University of Canterbury, Department of Physics and
Astronomy, Private Bag 4800, Christchurch 8020, New Zealand}

\affil{$^{5}$Korea University of Science and Technology, 
217 Gajeong-ro, Yuseong-gu, Daejeon 34113, Korea}

\affil{$^{6}$Department of Astronomy, Ohio State University, 140 W.
18th Ave., Columbus, OH 43210, USA}

\affil{$^{7}$Max-Planck-Institute for Astronomy, K\"{o}nigstuhl 17,
69117 Heidelberg, Germany}

\affil{$^{8}$Department of Physics, Chungbuk National University,
Cheongju 28644, Republic of Korea}

\affil{$^{9}$Harvard-Smithsonian Center for Astrophysics, 60 Garden Street, 
Cambridge, MA 02138, USA}

\affil{$^{10}$School of Space Research, Kyung Hee University,
Yongin, Kyeonggi 17104, Korea}

\affil{$^{11}$IPAC/NExScI, Mail Code 100-22, Caltech, 1200 E. 
California Blvd., Pasadena, CA 91125}

\affil{$^{12}$IPAC, Mail Code 100-22, Caltech, 1200 E. California Blvd., 
Pasadena, CA 91125}

\affil{$^{13}$Vanderbilt University, Department of Physics $\&$
Astronomy, Nashville, TN 37235, USA}

\affil{$^{\dag}$NASA Postdoctoral Program Fellow}

\affil{$^{\dag\dag}$Sagan Fellow}

\begin{abstract} 
We present microlensing planet OGLE-2017-BLG-0173Lb,
with planet-host mass ratio either $q\simeq 2.5\times 10^{-5}$ or
$q\simeq 6.5\times 10^{-5}$, the lowest or among the
lowest ever detected.  The planetary perturbation is
strongly detected, $\Delta\chi^2\sim 10,000$, because it
arises from a bright (therefore, large) source passing over and 
enveloping the planetary caustic: a so-called
``Hollywood'' event.  The factor $\sim 2.5$ offset in
$q$ arises because of a previously
unrecognized discrete degeneracy between Hollywood events in
which the caustic is fully enveloped and those in which only one flank
is enveloped, which we dub ``Cannae'' and ``von Schlieffen'',
respectively.  This degeneracy is ``accidental'' in that it
arises from gaps in the data.  Nevertheless, the
fact that it appears in a $\Delta\chi^2=10,000$ planetary anomaly is
striking.  We present a simple formalism to estimate
the sensitivity of other Hollywood events to planets and show that
they can lead to detections close to, but perhaps not quite reaching,
the Earth/Sun mass ratio of $3\times 10^{-6}$.  This formalism also
enables an analytic understanding of the factor $\sim 2.5$ offset in
$q$ between the Cannae and von Schlieffen solutions.  The 
Bayesian estimates for the host-mass, system distance,
and planet-host projected separation are
$M=0.39^{+0.40}_{-0.24}\,M_\odot$, $D_L=4.8^{+1.5}_{-1.8}\,\kpc$, and
$a_\perp=3.8\pm 1.6\,\au$.  The two estimates of the planet mass are
$m_p=3.3^{+3.8}_{-2.1}\,M_\oplus$ and $m_p=8^{+11}_{-6}\,M_\oplus$.
The measured lens-source relative proper motion $\mu=6\,\masyr$ will
permit imaging of the lens in about 15 years or at first light on
adaptive-optics imagers on next-generation telescopes.
These will allow to measure the host mass but probably cannot resolve the 
planet-host mass-ratio degeneracy.
\end{abstract}


\section{{Introduction}
\label{sec:intro}}

Planetary companions of microlensing hosts induce two classes
of caustic structures on the otherwise smooth magnification pattern
due to the host itself; ``central caustics'' that lie projected close
to the host and ``planetary caustics'' that are separated
from the host position by $(s - s^{-1})\theta_\e$, where $s\theta_\e$
is the planet-star angular separation and $\theta_\e$ is the angular Einstein
radius\footnote{If $s$ is sufficiently close to unity, then these
two caustics merge into a ``resonant caustic'', but for present
purposes, i.e., the regime of very low mass-ratio planets, 
this can be considered as a variant of ``central caustics''.}
\citep{mao91,gouldloeb,gaudi12}.

For planets that lie outside the Einstein ring $(s>1)$, each of the
two caustics may be regarded as due to the tidal shear induced
by the other body on its own symmetric gravitational field,
a regime that was first analyzed in a cosmological context by
\citet{cr2}, and then in a microlens-planet context by \citet{gouldloeb}.
Since the Einstein radius $\theta_\e$ of each is proportional
to the square root of the lens mass $M$,
\begin{equation}
\theta_\e\equiv \sqrt{\kappa M\pi_\rel};
\qquad
\kappa\equiv {4G\over c^2\au}\simeq 8.14{\mas\over M_\odot},
\label{eqn:thetaedef}
\end{equation}
and the shear scales directly with $M$, it immediately follows that
the size $w$ of the planetary caustics is larger than that of the
central caustics by
\begin{equation}
{w_{\rm planet}\over w_{\rm host}} \propto 
{M_{\rm host}\over M_{\rm planet}}{\theta_{\e,\rm planet}\over\theta_{\e,\rm host}} 
\propto \sqrt{M_{\rm host}\over M_{\rm planet}}\equiv q^{-1/2}.
\label{eqn:wrat}
\end{equation}
Here $\pi_\rel\equiv \au(D_L^{-1}-D_S^{-1})$ is the lens-source relative
parallax and $q$ is the planet-star mass ratio.

It follows directly from Equation~(\ref{eqn:wrat}) that random source
trajectories passing through the Einstein ring will intersect
planetary caustics much more often than central caustics and that
this disparity should grow stronger with decreasing mass ratio $q$.
Naively, this would seem to imply that the great majority
of microlensing planet detections should take place via planetary
caustics.  In fact, a recent compilation by \citet{ob160596} shows
that fewer than one third are detected through this channel.
While examination of their Figure~7 does show that this fraction rises
for low-mass planets $q<2\times 10^{-4}$ (as one would expect from
the above argument), it is still barely more than 50\% in this low-mass
regime.

The main reason for this discrepancy is that while sources do
pass randomly through Einstein rings, they are not all monitored
equally.  \citet{griest98} and \citet{rattenbury02}
showed that high-magnification events
are much more sensitive to planets simply because (by definition)
the source trajectory goes very close to the host, where every planet
induces distortions in the magnification profile via a central caustic.
For this reason, if there are limited observational resources, and
if the high-magnification events can be recognized in time to
mobilize these resources, it makes sense to concentrate them
on high-magnification events.  See, for example, \citet{gould10}.

Wide-area high-cadence surveys, initiated first by the Microlensing
Observations in Astrophysics (MOA, \citealt{mb13685}) 
and the Optical Gravitational Lensing
Experiment (OGLE, \citealt{ogleref}) 
can continuously monitor all events, whether the
source trajectories are individually favorable or not, making them
much more sensitive to planetary caustics and therefore, in particular 
according to Equation~(\ref{eqn:wrat}), low-mass planets.  Indeed, this
is a major component of the motivation for such surveys.

For gas-giant planets, whose characteristic Einstein timescales
(and so typical durations of perturbation)
are one-to-few days, the sensitivity of the survey is basically
independent of the diurnal coverage.  For example, OGLE-2012-BLG-0406Lb,
one of the first planets to be detected in this mode, has a perturbation
spanning about five days.  OGLE data, which are taken from Chile, quite 
adequately
cover the anomaly, enabling robust characterization, even though
they span only about 1/3 of the diurnal cycle and even though
three days were entirely missed due to the Moon passing through
the Galactic bulge \citep{ob120406}.  These results were later
confirmed and refined by \citet{ob120406b}.

For planets at the opposite extreme, i.e., Earth-to-Neptune mass-ratio
planets with characteristic timescales of a few hours, these surveys
retain their sensitivity, but it now scales as their diurnal coverage.
That is, such short-duration anomalies are likely to be either
basically contained within a night's data or entirely missed.

Detection of such low-mass planets is a major motivation for
creating round-the-clock surveys, either by combining several
surveys located at complementary sites \citep{mb11322}, or by
organizing a single, multi-site survey \citep{kmtnet,henderson14}.

Here we report the discovery of OGLE-2017-BLG-0173Lb, which 
at either $q\simeq 2.5\times 10^{-5}$ or
$q\simeq6.5\times 10^{-5}$, is in this latter regime:
the discovery relies critically on the near-continuous
coverage of the Korea Microlensing Telescope Network (KMTNet) survey,
with the anomaly entirely captured by data taken from its Australia and
South African observatories.

The source star of the event is a giant and so has a large angular
radius.  The event therefore illustrates the power of the so-called
``Hollywood'' strategy of ``following the big stars'' to detect
low mass planets \citep{gould97}.

\section{{Observations}
\label{sec:obs}}

OGLE-2017-BLG-0173 is at (RA,Dec) = (17:51:52.95,$-29$:16:16.9),
corresponding to $(l,b)=(0.42,-1.35)$.  
It was discovered and announced as a probable microlensing event
by the OGLE Early Warning
System \citep{ews1,ews2} at UT 14:21 25 Feb 2017.  OGLE observations
were at a cadence of $\Gamma = 1\,{\rm hr^{-1}}$ using their 1.3m telescope
at Las Campanas, Chile.

KMTNet observed this field from its three 1.6m
telescopes at CTIO (Chile, KMTC), SAAO (South Africa, KMTS) and SSO 
(Australia, KMTA),
in its two slightly offset fields BLG02 and BLG42, with combined
cadence of $\Gamma = 4\,{\rm hr^{-1}}$.  However, for KMTC-BLG02,
the source usually fell on a bad column of the detector, so observations
from this observatory-field combination were not included.  Hence,
For KMTC, $\Gamma = 2\,{\rm hr^{-1}}$.  

The great majority of observations were carried out in $I$ band 
with occasional $V$ band observations made
solely to determine source colors (but see next paragraph).  
All reductions for the light curve
analysis were conducted using variants of difference image analysis (DIA,
\citealt{alard98}), specifically \citet{wozniak2000} and \citet{albrow09}.

\begin{figure}
\plotone{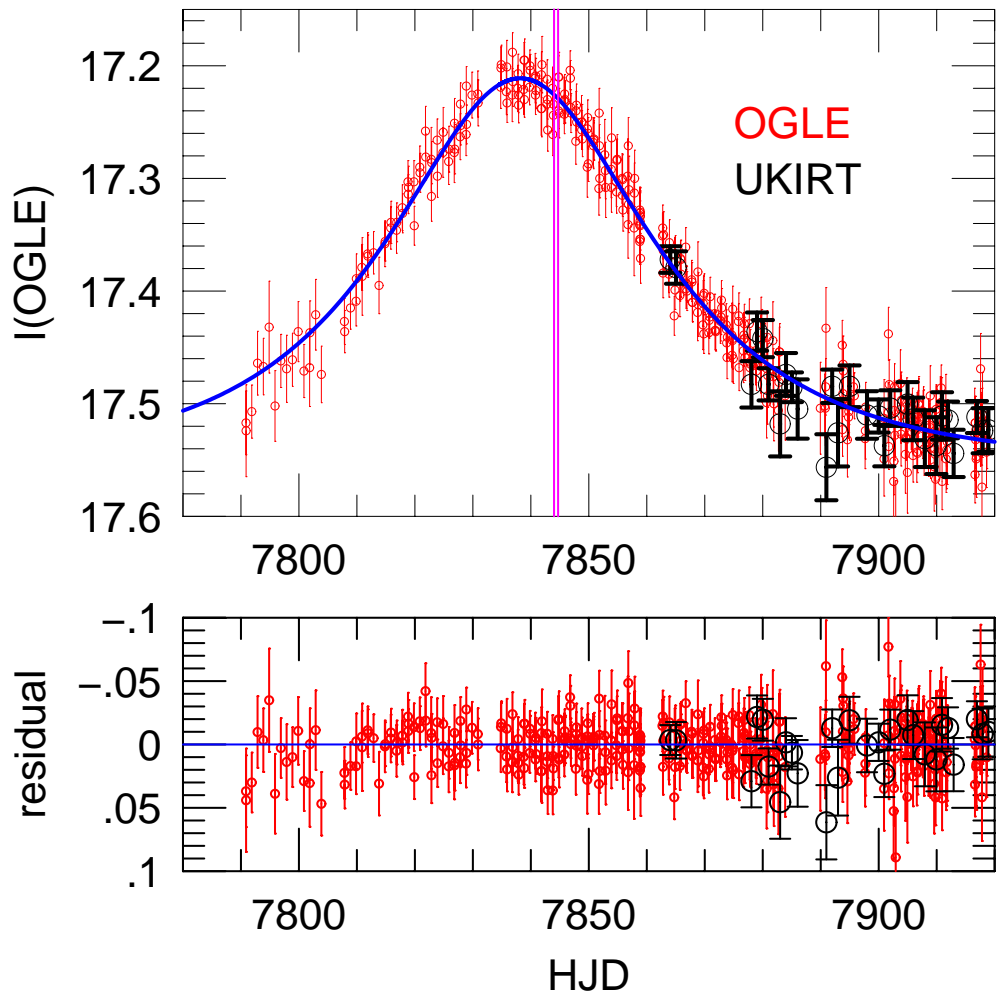}
\caption{OGLE and UKIRT data for OGLE-2017-BLG-0173.  The OGLE data (red)
trace a seemingly normal low-amplitude microlensing event.  The
time interval of the anomaly, marked by a pair of magenta vertical lines,
by chance does not overlap the periods of visibility from Chile.
The black points show the UKIRT $H$-band data (transformed to the OGLE
scale).  Although these begin well after peak, they enable an $(I-H)$ color
measurement.  
}
\label{fig:ogleukirt}
\end{figure}

Although the source is a low-luminosity giant, and therefore quite
luminous relative to the majority of microlensed sources, it is also
highly extincted, $A_I=2.8$.  Hence, it is extremely faint in $V$ band.
While many faint-$V$ sources nevertheless can ultimately yield very good
$(V-I)$ colors (e.g., \citealt{mb11293}), this is only because they
are observed at high magnification.  By contrast, the most highly
magnified $V$ point for OGLE-2017-BLG-0173
has a magnification $A\sim 2.2$ (i.e., $\Delta A=1.2$
relative to baseline), which yields only very poor constraints
on the source color.  
Fortunately, the UKIRT microlensing survey \citep{Shvartzvald.2017.A}, 
which is primarily motivated to improve understanding of future 
{\it WFIRST} microlensing observations \citep{spergel13},
observed this field using the wide-field near infrared camera 
(WFCAM) with a nominal cadence of $\Gamma=1\,{\rm day}^{-1}$ in $H$-band.
Although these observations began 26 days after peak, when the source was
just leaving the Einstein ring (so magnified by only 
$\Delta A\sim 0.3$, see Figure~\ref{fig:ogleukirt}), the
source was quite bright in this passband, $H_s\sim 14.5$, which enables
a good $(I-H)$ color measurement.  See Section~\ref{sec:cmd}.
The UKIRT/WFCAM images were reduced by the Cambridge Astronomy Survey Unit
(CASU; \citealt{Irwin.2004.A}).
The UKIRT light curve was extracted using a soft-edged circular aperture and
was photometrically calibrated to 2MASS (see \citealt{Hodgkin.2009.A} for
details).

\section{{Analysis}
\label{sec:analysis}}

The OGLE-2017-BLG-0173 light curve is comprised of a long low-amplitude
hump, lasting several months which rises just 0.35 mag above baseline
(see Figure~\ref{fig:ogleukirt}), 
punctuated by a short $(<1\,{\rm day})$ bump, which rises an additional
0.3 mag (see Figure~\ref{fig:lc}).  The anomaly starts and ends
with an abrupt rise and fall, indicating a caustic entrance and exit
respectively.  However, there is no ``dip'' between these, implying
that the source must be larger than the separation between the two sides
of the caustic.  The simplest way to account for this behavior is that
the source envelops a substantial fraction of, or perhaps the whole, caustic.
While a few such events have previously been observed,
e.g., OGLE-2005-BLG-390 \citep{ob05390} and OGLE-2008-BLG-092
\citep{ob08092}, there has never been a discussion of how to
intuitively understand this generic class of events.  We therefore
begin with such a heuristic analysis.

\subsection{{Heuristic Analysis}
\label{sec:heuristic}}

In general, a minimum of six geometric parameters are required
to describe a binary microlensing event: $(t_0,u_0,t_\e,s,q,\alpha)$.
The first three are the \citet{pac86} parameters of the underlying
event due to the system as a whole, 
i.e., the time of closest approach, the impact
parameter (scaled to $\theta_\e$) and the Einstein timescale\textbf{,}
\begin{equation}
t_\e\equiv {\theta_\e\over\mu},
\label{eqn:tedef}
\end{equation}
where $\bmu$ is the lens-source relative proper motion and $\mu=|\bmu|$.  
For the
case of planetary companions, particularly those of very low mass,
the great majority of the light curve follows the standard \citet{pac86}
flux evolution\textbf{:}
\begin{equation}
F(t) = f_s A[u(t;t_0,u_0,t_\e)] + f_b;
\qquad
A(u) = {u^2 + 2\over u\sqrt{u^2+4}};
\qquad
u^2 = {(t-t_0)^2\over t_\e^2} + u_0^2,
\label{eqn:pac86}
\end{equation}
where $f_s$ is the source flux and $f_b$ is any blended light
in the aperture not taking part in the event.  The normalized
separation $s$ and mass ratio $q$ have already been described,
while $\alpha$ gives the direction of the star-planet axis relative to $\bmu$.
If the source passes over or close to any caustics, one must also
specify a seventh parameter\textbf{,}
\begin{equation}
\rho\equiv {\theta_*\over\theta_\e},
\label{eqn:rhoef}
\end{equation}
where $\theta_*$ is the angular source radius.

As shown in Figure~\ref{fig:ogleukirt}, the OGLE data are quite well
fit by a standard \citet{pac86} curve, Equation~(\ref{eqn:pac86}).
After including the KMTNet data as well (but excluding the anomaly)
we find that the \citet{pac86} parameters are,
\begin{equation}
(t_0,u_0,t_\e,f_s,f_b)=(7837.90,  0.74, 33.7\,{\rm day},  0.93,  0.61),
\qquad ({\rm point-lens\ fit}),
\label{eqn:pac86eval}
\end{equation}
where the flux scale is set by $f\equiv 1$ at $I=18$.  In addition
to these five parameters, we must still evaluate the four others 
related to the planet that were defined above,
$(s,\alpha,q,\rho)$.  The first two of these are quite straightforward.

The perturbation is centered at $t_{\rm anom}=7844.3$, i.e., at
$\tau_{\rm anom}\equiv (t_{\rm anom}-t_0)/t_\e = 0.190$.  Hence, the position
and orientation within the Einstein ring are
\begin{equation}
u_{\rm anom} = \sqrt{u_0^2+\tau_{\rm anom}^2} = 0.76  
\qquad
\alpha =\tan^{-1}{u_0\over\tau_{\rm anom}}= 1.32\,{\rm radian}. 
\label{eqn:alpha+upert}
\end{equation}
The physical origin of the caustic is that one of the two images created
by the gravity of the host,
at scaled positions $(u\pm \sqrt{u^2+4})/2$, is passing near the planet,
with separation $s$ from the host.  However, as shown by \citet{gg97},
when a source envelops the caustics due to the ``minor'' image
($(u- \sqrt{u^2+4})/2$), it tends to generate zero excess magnification
rather than the bump that is seen in Figure~\ref{fig:lc}.  Hence
we derive,
\begin{equation}
\qquad s = {u_{\rm anom} + \sqrt{u_{\rm anom}+4}\over 2} = 1.45\ . 
\label{eqn:seval}
\end{equation}

To evaluate the remaining two parameters $(q,\rho)$ by eye is more
difficult.  We begin by making the simplifying assumption
that the source completely envelops the caustic and then discuss
how the estimates are impacted if this assumption fails and the 
caustic is only partially enveloped.

\citet{gg97} showed that for the case of an $s>1$ planetary caustic, 
the excess magnification at peak for a source that is much larger
than the planetary Einstein radius $\theta_{\e,p}\equiv \sqrt{q}\theta_\e$ is
\begin{equation}
\Delta A = {2q\over\rho^2}.
\label{eqn:gg}
\end{equation}
That is, $\Delta A$ is the same as it would be if the planet were
an isolated point-lens.

The excess flux can be read off the light curve, which then
(combined with $f_s$) yields the excess magnification\textbf{,}
\begin{equation}
\Delta A= {10^{-0.4\,I_{\rm anom,peak}} - 10^{-0.4\,I_{\rm anom,base}}
\over 10^{-0.4\,I_s}}=0.67;  
\qquad
{q\over\rho^2} = {\Delta A\over 2} = 0.33, 
\label{eqn:deltaa}
\end{equation}
where $I_{\rm anom,peak}=16.94$ and $I_{\rm anom,base}=17.23$\ .

In order to evaluate $q$, we must determine $\rho$.  
Working in the limit that the source is much larger than the
planetary Einstein radius ($q/\rho^2\ll 1$), which is only marginally
satisfied by Equation~(\ref{eqn:deltaa}), and assuming that 
the source center passes directly over the caustic, then
$\rho = t_{\rm fwhm}/2t_\e$, where $t_{\rm fwhm} = 0.5\,$days, is the
full width at half maximum of the bump.  Under this
assumption\textbf{,}
\begin{equation}
\rho = {t_{\rm fwhm}\over 2 t_\e} = 0.0074; 
\qquad
q = {\Delta A\over 2}\rho^2 = {t_{\rm fwhm}^2\Delta A\over 8 t_\e^2} = 1.84\times
10^{-5}.  
\label{eqn:rhoeval}
\end{equation}

The above formalism is appropriate if the source fully envelops
the caustic, which we dub ``Cannae'' events.  However, qualitatively
similar event morphologies will be generated if the source envelops
only one flank of the caustic, which we call ``von Schlieffen'' events.
As a representative of these, we consider the case
that the limb of the source (rather than its center)
passes directly over the center of the caustic.  Then, the above
argument gives us $q=\Delta A\rho^2$, which yields $q=3.7\times 10^{-5}$. 

Hence, the heuristic analysis indicates that $10^{-5}\la q \la 10^{-4}$,
but more detailed numerical analysis is needed to make a more precise
estimate.  More generally, this analysis tells us that planets of
quite small mass ratio are easily detectable in these large-source
events.

\subsection{Numerical Analysis}
\label{sec:numerical}

To carry out a systematic analysis, we begin (as described above) by fitting a
\citet{pac86} curve to the full data set excluding the anomaly, in
order to obtain initial estimates of $(t_0,u_0,t_\e)$.  We also
make an initial estimate of $\rho=0.01$ following the reasoning above.
We then conduct a grid search over $(s,q)$ space, holding these two parameters 
fixed and allowing the other five to vary, including $\alpha$
which we seed at a grid of values.  We use inverse-ray shooting 
\citep{kayser86,schneider88,wambs97} when the source is close to a
caustic and multipole approximations \citep{pejcha09,gould08}
otherwise.  We employ Monte Carlo Markov Chain (MCMC) to locate all minima.
We find three solutions (A,B,C).  All three have very similar geometries
defined by $(t_0,u_0,t_\e,\rho,s,\alpha)$, but one of them (B) has 
a Cannae topology and the other two have von Schlieffen topologies,
one on each flank.  Note that the degeneracy between solutions A and C
was already predicted by \citet{gaudi97}, but the degeneracy of these two
solutions with B has not previously been predicted nor seen in practice.
Figure~\ref{fig:lc} shows the three model light curves
superposed on the data, while Figure~\ref{fig:geom} shows the lens-source
geometries and the resulting relations between the source and the caustics.
The best fit parameters and uncertainties are shown in Table~\ref{tab:ulens}.  
Note that all of the parameters are in reasonable agreement with
those derived from the heuristic analysis of 
Section~\ref{sec:heuristic}.   In particular, for the Cannae solution,
the estimated mass ratio $q$ is too low by about 25\% and for the
von Schlieffen solutions it is too low by about 40\%.

As demonstrated by Table~\ref{tab:ulens}, the three solutions are 
extremely close in terms of $(t_0,u_0,s,\alpha)$ geometry, but are 
strongly separated in $q$.  We find that the three minima are discrete
in the sense that the MCMC does not jump from one to the other in
a normal run, showing that the barriers between them are too high.
To explore the nature of these barriers, we run a ``hotter'' MCMC
(artificially inflating the error bars by a factor 3.0 and then,
to compensate for this, multiplying the resulting $\chi^2$ values
by 9.0).  To trace the relation between these solutions, we introduce
the parameter
\begin{equation}
\Delta\xi \equiv u_0\csc\alpha - \xi_+(s);
\qquad
\xi_+(s)\equiv s - s^{-1},
\label{eqn:deltaxi}
\end{equation}
where $[\xi_+(s),0]$ is the Einstein-ring position of the center
of the major-image caustic for a planet in the regime $q\ll 1$
with separation $s$.  That is, $\Delta\xi$ is the offset between
the centers of the source and the caustic as the source 
crosses planet-star axis.

Figure~\ref{fig:deltaxi} plots $\log q$ vs.\ $\Delta\xi$.  
It shows a broad minimum in $q$ centered near $\Delta\xi=0$ (solution B), with
$q$ rising roughly symmetrically toward solutions A and C on either
side.  These tracks are continuous in the parameters of the plot,
but have relatively high barriers between the three minima, as expected.
We find that these barriers have heights of $\Delta\chi^2\sim 35$
between solutions A and B, and $\Delta\chi^2\sim 60$
between solutions B and C.

The $\chi^2$ differences between the best (B) and worst (C) solutions
is $\Delta\chi^2\sim 16$, and hence solution C can be considered
as strongly disfavored.  However, the two von Schlieffen
solutions (A and C) have almost identical physical implications, 
while solutions A and B have very similar $\chi^2$.  Hence, the degeneracy
between solutions with different $q$ (and so different physical implications)
is quite severe.

\subsection{Reality of Roughly Equal Source and Blend Fluxes?}
\label{sec:blendquest}

A mildly peculiar feature of these solutions is that $f_s$ and $f_b$
are comparable.  This is more true of solution C than either A or B.
Nevertheless, this statement qualitatively
describes all three solutions.
Such rough equality is frequently observed for
typical microlensed sources, which are most often stars near the
turnoff.  Since the projected density of stars of similar luminosity
is very high toward the bulge, it is not at all uncommon to have more
than one in a ground-based seeing disk.  However, as we will discuss
in Section~\ref{sec:cmd}, the ``baseline object'' is in or near
the Galactic bulge clump, and it would be much rarer to have a star
of comparable brightness projected on such a source.  The issue is
important because spurious blending could be an indication of
systematic errors in the data that are driving the solution.

Fortunately, we have very clear evidence that the blending is real,
which also permits us to estimate the allowed range of $f_b/f_s$
completely independent of the light curve analysis.

We find that the astrometric position of the ``baseline object''
(i.e., the cataloged ``star'' derived from analysis of the
template image) is offset from the position of the source by 
$\Delta\theta_{\rm app}=0.13^{\prime\prime}$.  The source position is derived by
finding the position of the ``difference star'' formed by subtracting
the template image from images taken near peak magnification.  The
offset between the blending star (or the light centroid of several
blending stars) and the source, $\Delta\theta$, is related to the
apparent offset by
\begin{equation}
\Delta\theta_{\rm app} = {f_b\Delta\theta\over f_s + f_b}
\qquad\Longrightarrow\qquad {f_b\over f_s} = 
\biggl({\Delta\theta\over\Delta\theta_{\rm app}}-1\biggr)^{-1}
\label{eqn:dtheta}
\end{equation}
We consider that if the blend were separated by more than one FWHM
in the very good seeing images of the template,  FWHM$\sim 0.8^{\prime\prime}$,
then it would have been separately resolved, i.e., 
$\Delta\theta\la 0.8^{\prime\prime}$.  Hence, Equation~(\ref{eqn:dtheta})
implies: $0.19\la f_b/f_s<\infty$.  From Table~\ref{tab:ulens}, the
best fit values for this ratio are $f_b/f_s=(0.34,0.28,0.60)$ for
solutions (A,B,C), which are all easily satisfied.  Alternatively,
we can derive best fit estimates 
$\Delta\theta=(0.51^{\prime\prime},0.59^{\prime\prime},0.35^{\prime\prime})$.
In all cases it is quite plausible that blends with the corresponding
flux ratios would not be detected at these separations in the template.

\subsection{Binary-Source Solution?}
\label{sec:binarysource}

As pointed out by \citet{gaudi98} short-term peaks that are the hallmark
of planetary perturbations can also be generated by a second source,
which then typically should pass very close to the lens (accounting for its
short apparent timescale) and would then also be 
very faint (so as not to completely
dominate the light curve during this close passage).  We search for such
solutions but do not find acceptable fits.  See Figure~\ref{fig:lc}.
The basic reason for this is that the perturbation is simply too compact.  

To better understand the underlying reasons for this numerical result,
we first consider the case that the second source is not impacted by
finite source effects.  Then the effective timescale 
$t_{\eff,2} \equiv u_{0,2}t_\e \simeq t_{\rm fwhm}/\sqrt{12} = 0.14\,{\rm day}$,
which implies that the excess flux during the Chile observations at 
$\Delta t=0.45\,{\rm day}$ would be 
$f_{\rm exc} = f_{2,\rm peak}/\sqrt{1+(\Delta t/t_\eff)^2} 
\simeq 12^{-1/2}(t_{\rm fwhm}/\Delta t)f_{2,\rm peak}
= 0.2$, corresponding
to a change of 0.1 magnitudes.  This would clearly contradict the data.

On the other hand, consider the case that the second source passes
directly over the lens.  One may show\footnote{
See, e.g., Figure 3 of \citet{ob151482} and note that
$t_{\rm fwhm} = 2 z_{1/2}\rho t_\e$, where $z_{1/2}\simeq 1.1$ is the
solution of $B(z_{1/2})/z_{1/2} = (1/2)B^\prime(0) = 1$.} that
$t_{\rm fwhm}\simeq 2.2\,\rho t_\e$.  Then $f_{s,2} = f_{2.\rm peak}\rho_2/2$, 
and hence the excess flux seen from Chile just before the bump would be 
$f_{\rm exc} = f_{s,2}t_\e/\Delta t = f_{2,\rm peak} t_{\rm fwhm}/4.4\Delta t= 0.17$,
corresponding to about 0.08 mag.  This is still clearly excluded by the data.


\section{Physical Parameters}
\label{sec:phys}

The lens mass $M$ and lens-source relative parallax $\pi_\rel$ can
in principle be determined provided that both the Einstein
radius $\theta_\e$ (Equation~(\ref{eqn:thetaedef})) and the microlens
parallax \citep{gould92,gould00,gouldhorne},
\begin{equation}
\bpi_\e \equiv {\pi_\rel\over\theta_\e}\,{\bmu\over\mu},
\label{eqn:piedef}
\end{equation}
can be measured.  Then, $M=\theta_\e/\kappa\pi_\e$ and 
$\pi_\rel=\theta_\e\pi_\e$.  As in most planetary microlensing events,
$\theta_\e$ can be measured, but unfortunately we find that $\bpi_\e$
can be neither measured nor meaningfully constrained. Therefore,
after measuring $\theta_\e$ in Section~\ref{sec:cmd}, 
we apply a Galactic model to estimate the lens mass and distance.
\subsection{Measurement of $\theta_\e$ and $\mu$}
\label{sec:cmd}

As mentioned in Section~\ref{sec:obs}, we are able to roughly place
the ``baseline object'' on a $[(V-I),I]$ color-magnitude diagram (CMD),
but we cannot actually measure the  source color in these bands.
This is due partly to its faintness in $V$ (as a result of
the $A_V\sim 5$ magnitudes of extinction), and partly because
its peak magnification is quite modest.  We therefore use UKIRT
$H$ band in place of the usual $V$ band, to determine the color.

As illustrated in Figure~\ref{fig:cmd}, the source is
$0.16\pm 0.06$ mag redder and for solutions (A,B,C) 
$(0.60,0.55,0.79)\pm (0.09,0.11,0.07)$
mag fainter than the clump
in these bands.  
We use \citet{bb88} to convert the color offset to
$\Delta(V-I)=0.13\pm 0.06$. 
Then adopting
$[(V-I),I]_{0,\rm clump}=(1.06,14.43)$ from \citet{bensby13} and
\citet{nataf13}, we find $[(V-I),I]_{0,s}=[1.19,(15.03,14.98,15.22)]$.  
Again using
the color-color relations of \citet{bb88} as well as the
the color/surface-brightness relation of \citet{kervella04}, we find,
\begin{equation}
\theta_* = (5.30,5.43,4.86)\pm (0.30,0.35,0.26)\,\muas ,
\label{eqn:thetastar}
\end{equation}
and so (taking account of the correlation between $\rho$ and $f_s$),
\begin{equation}
\theta_\e ={\theta_*\over\rho} = (0.48,0.54,0.53)\pm 0.03\,\mas;
\quad
\mu ={\theta_\e\over t_\e} = (5.7,6.5,5.9)\pm (0.3,0.4,0.3)\,\masyr .
\label{eqn:thetaemu}
\end{equation}

\subsection{Bayesian Estimate}
\label{sec:bayes}

The angular Einstein radius $\theta_\e$ and relative proper motion $\mu$
(Equation~(\ref{eqn:thetaemu}))
are quite consistent with either a disk or bulge lens.
See upper left panel of Figure~7 of \citet{penny16}.
To make a more quantitative estimate of the lens characteristics,
we draw lensing events randomly from a \citet{han95,han03} Galactic model 
and catalog
the subset that are consistent with the observables, $\theta_\e$
and $\mu$.  

The results are shown in Table~\ref{tab:phys}.  For host mass, system distance,
and planet-host projected separation, the results are essentially the same
for the three solutions:
$M=0.39^{+0.40}_{-0.24}\,M_\odot$, $D_L=4.8^{+1.5}_{-1.8}\,\kpc$, and
$a_\perp=3.8\pm 1.6\,\au$.  
However, since $q$ is very different for the von Schlieffen solutions
compared to the Cannae solution, the planet masses are also centered at
very different values, $m_p=8\,M_\oplus$ and $m_P=3.3\,M_\oplus$, respectively.
Since the fractional error in $q$ is much smaller than that of the Bayesian
estimate of $M$, the fractional error in $m_p$ is dominated by the latter.

We show the posterior histograms for $M$ and $D_L$ for solution B in
Figure~\ref{fig:hist}.  The corresponding figures for the other two
solutions are extremely similar and so are not shown.

\subsection{Future Resolution}
\label{sec:resolve}

Because of its low mass ratio, either $q\simeq 6.5\times 10^{-5}$ 
(solutions A and C) or $q\simeq 2.5\times 10^{-5}$ (solution B), it would
be of significant interest to determine the true mass of the host
and to resolve the degeneracies among the three solutions, and thereby
determine the mass of the planet.  Here we show that the first will eventually
be possible and that the second will probably not be possible.

If the event had occurred somewhat later in the season, it could
have been targeted for {\it Spitzer} microlensing observations
\citep{prop2016}.  However, the first epoch at which
it was visible by {\it Spitzer} 
was at HJD$^\prime=7930$ when $u\sim 2.7$, and
hence $A\sim 1.024$, i.e., quite close to baseline.  Such
observations were nevertheless attempted, but did not
yield useful constraints.

Hence, the best hope for measuring the host mass is to image
the lens when it has sufficiently separated from the source,
using high-resolution imaging.  Since the source is likely
to be 100--1000 times brighter than the lens, it will probably
require (with current instruments) about 1.5 times larger
separation than the $60\,\mas$ separation by which \citet{ob05169bat}
resolved the roughly equal-brightness lens and source of OGLE-2005-BLG-169.
 From the measured proper motion, this would require a roughly 15 year
wait. In the meantime, high-resolution imagers of next generation
(``30 meter'') telescopes may come on line, in which case the
lens could be imaged at first light.

To aid with these measurements, which may be decades in the
future, we give a short summary of what is known from the event about
the $H$-band fluxes and their errors together with the underlying
reason.  The most precise measurement is of the $(I-H)$ source
color, where $I$ is in the OGLE-IV system and $H$ is in the 2MASS system:
$(I-H)_s = 3.71\pm 0.06$.  This is essentially independent of any
model and depends on regression and the assumption of achromaticity,
which follows from general relativity. At the next level, we have
$H_s = I_s - (I-H)_s = (14.14,14.09,14.34)\pm (0.11,0.12,0.09)$ for (A,B,C),
where $(I-H)_s$ is described just
above and the $I_s$ are derived from the models (see Table~\ref{tab:ulens}).
Since the $I_s$ and $(I-H)_s$ measurements are essentially 
independent, the errors
are added in quadrature.  Finally, we report the $H$-band baseline
$H_{\rm base} = 14.01 \pm 0.01$.  Because there are enough data
very close to baseline, the error in this quantity is basically
just the calibration error.  Therefore, subtracting fluxes and
noting that the $H_b$ and $H_s$ flux errors are the same, we
obtain $H_b=(16.37,16.88,15.46)\pm (0.86,1.56,0.25)$.

By directly resolving the host and measuring both its color and magnitude
(and combining this with the mass-distance constraint from the
measurement of $\theta_\e = \sqrt{\kappa M\pi_\rel}$), the host mass can be
determined.  

Unfortunately, such a measurement would not discriminate among
the three solutions.  All three predict similar lens masses and distances.
This is particularly true of solution A and B,
which differ by only $\Delta\chi^2=2.5$.

However, the same high-resolution imaging (or even a high resolution
image taken much sooner) could in principle partially discriminate
among solutions by separately resolving the source and  the blended light.  
Suppose, for example, that the blended light were measured to have
$I_b=19.0$, corresponding to $f_b=0.40$.  From Table~\ref{tab:ulens},
this would be consistent with solutions A and B, but inconsistent
with solution C.  Unfortunately, solutions A and B predict the
same $f_b$ to well within $1\,\sigma$ so there is no possibility
of distinguishing between them by measuring $f_b$.

\section{Discussion}
\label{sec:discuss}

\subsection{Nature of Degeneracy}
\label{sec:degen}

As just discussed in Section~\ref{sec:resolve}, the factor 2.5 degeneracy
in the planet-host mass ratio between solutions A+C and solution B
cannot be resolved from the existing data and may never be resolved.  Hence,
while there is no doubt that OGLE-2017-BLG-0173Lb is a very low 
mass-ratio planet $q<10^{-4}$, and may be the lowest yet detected
$(q\simeq 2.5\times 10^{-5})$ its actual mass ratio remains
somewhat uncertain.  Given that the planetary deviation is detected
with extremely high confidence $(\Delta\chi^2\sim 10,000)$, it is of
considerable interest to understand the nature of the degeneracy.

Inspection of Figure~\ref{fig:lc} shows that this is an ``accidental''
degeneracy in that the model light curves differ significantly in
regions of the anomaly where there are gaps in the data.  These
in turn are due to the fact the anomaly occurred quite early in the
season (31 March), when KMTNet was able to observe only 4.0 hours
from each site.  In particular, models A and B are well separated
during the $\sim 3.5\,$hrs prior to the onset of KMTA observations.
While models B and C are less well separated, model C is already
disfavored by $\Delta\chi^2=16$.  Hence, it is likely that this
degeneracy would have been resolved if the event had occurred,
for example, 2 months later.

\subsection{Poster Child for ``Hollywood'' Events}
\label{sec:hollywood}

The mass ratio, $q\simeq 2.5\times 10^{-5}$ or $q\simeq 6.5\times 10^{-5}$,
of OGLE-2017-BLG-0173 is among the 
lowest for any microlensing planet (see e.g., Figure~7 from
\citealt{ob160596}).  Yet the
signal is quite strong.  From the analysis given in 
Section~\ref{sec:heuristic}, if $q$ had been substantially smaller,
say $q=10^{-5}$, (and focusing for the moment on solution B),
then the light curve would have looked qualitatively similar
but with the amplitude of the bump, $\Delta A$, reduced by a factor
of 2.7.  In this case, it still
would have been easily recognized.  Moreover,
the excess magnification would have been the same, regardless of 
the planet-host separation, provided $s>1$ and of course provided
that the source passed over the caustic.  For example, \citet{ob08092}
discovered a $q=2.4\times 10^{-4}$ planet from the passage of a giant source
over an $s=5.26$ planetary caustic, with $\rho=0.04$, in OGLE-2008-BLG-092.
Hence, $\Delta A = 2q/\rho^2=0.3$, which is quite similar to the present case.
In addition, \citet{mb12006} found a $q=0.016$ companion from a giant source
(very similar to the OGLE-2017-BLG-0173 source, with similar $\rho\sim 0.01$)
passing over an outlying caustic, $s=4.4$, in MOA-2012-BLG-006.
However, in this case, the large mass ratio implied that the companion
Einstein radius was about 10 times larger than the source, so that
$\Delta A\gg 1$, and hence the framework of Section~\ref{sec:heuristic}
does not apply.

\begin{figure}
\plotone{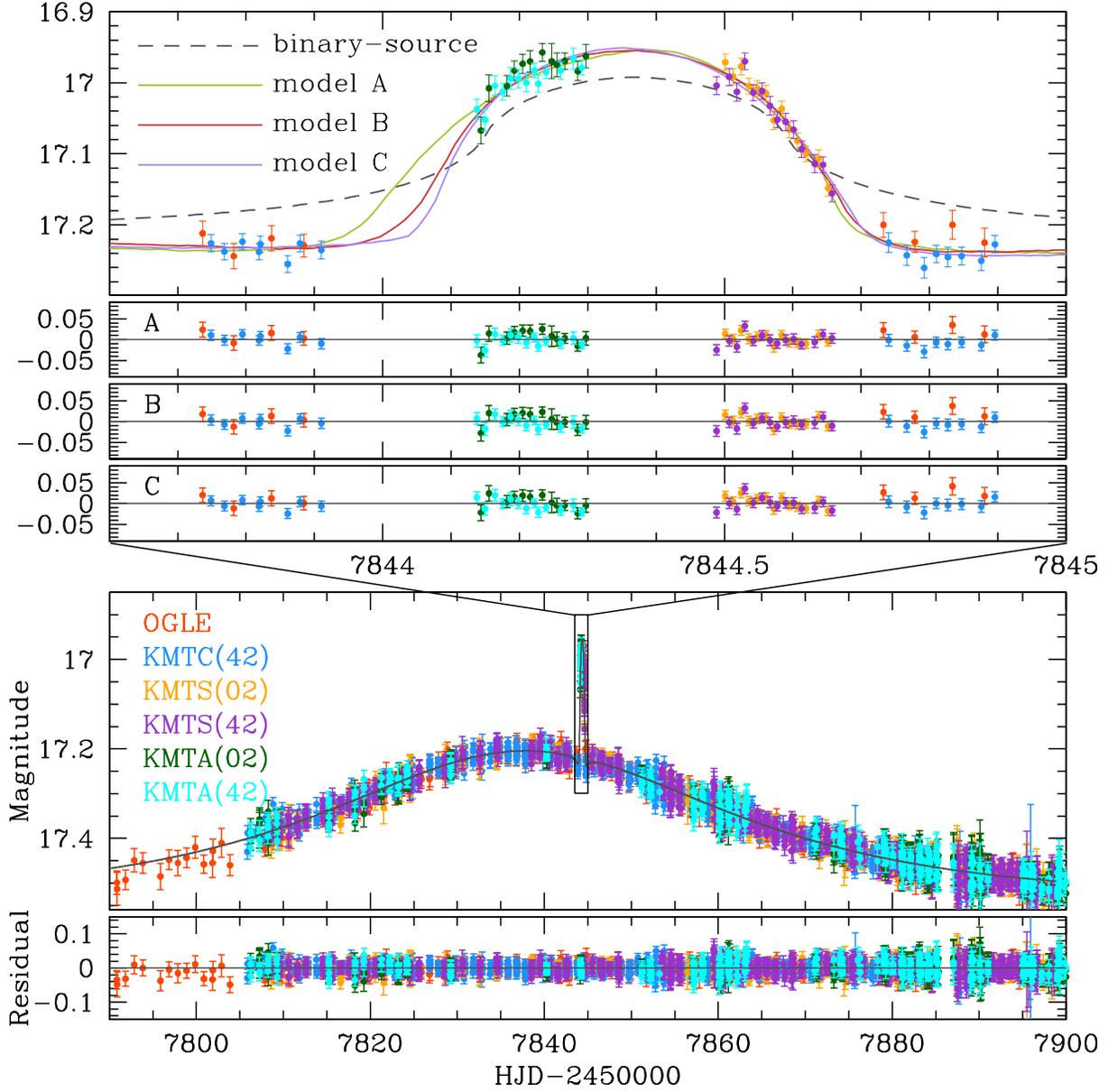}
\caption{Light curve and models of OGLE-2017-BLG-0173.  The short anomaly
at HJD$^\prime=7844.3$ on an otherwise perfectly normal point-lens
\citet{pac86} curve is due to a planet with mass ratio either
$q\simeq 2.5\times 10^{-5}$ (model B) or 
$q\simeq 6.5\times 10^{-5}$ (models A and C).  The upper panel 
is a zoom and also includes the best-fit 
binary-source model, which clearly fails to account for the data.
}
\label{fig:lc}
\end{figure}

\begin{figure}
\plotone{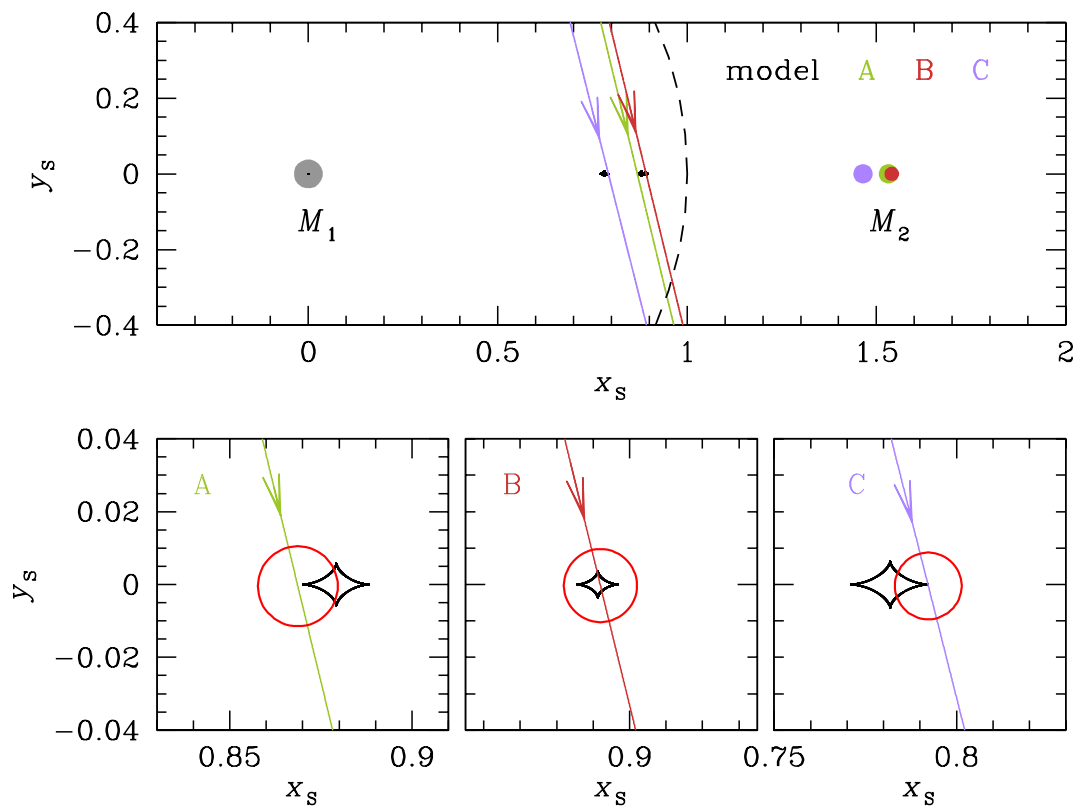}
\caption{Geometries of the three models (A,B,C) of OGLE-2017-BLG-0173.
The upper panel shows source trajectory and lens-component positions
color coded by model as well as the location of the caustic.  The
zooms in the lower panel show that
the giant-star source either partially (A,C) or fully (B) envelops
the caustic, making this either a ``von Schlieffen'' 
or ``Cannae'' type ``Hollywood'' event. 
}
\label{fig:geom}
\end{figure}

These facts illustrate the strengths
of the ``Hollywood strategy'' advocated by \citet{gould97} of
``following the big stars'' to find 
planets\footnote{Originally, the nickname ``Hollywood'' developed 
because cases in which
the star is big enough that it can envelop the caustic generally
correspond to cases in which the star is a giant and therefore are
often the brightest objects in the field. However, we use the term
here more generally to describe any case in which the source is
comparable to or larger than the caustic.  In fact, in the
{\it WFIRST} era \citep{spergel13}, even dwarf stars may fully
envelop the tiny caustics of the extremely low-mass planets
that will be detectable.}.
Whenever the star
is so big that it can envelop the caustic, the cross section for
anomalous deviations from a \citet{pac86} curve grows from
the size of the caustic (or a bit more) to the size of the source.
The duration of the deviation likewise grows, implying that 
modest-duration breaks in the light-curve coverage (like
the ones before and after KMTA observations near HJD$^\prime=7844$
in Figure~\ref{fig:lc}) do not compromise the detection
(although, as discussed in Section~\ref{sec:degen}, they can
degrade  characterization).

To highlight these points, we show in Figures~\ref{fig:artificialA},
\ref{fig:artificialB} and ~\ref{fig:artificialC},
simulated events that are geometrically identical to the real
one for models (A,B,C), except with a source that is 9.6 times smaller 
than the real source.  For didactic purposes, the top panel 
in each Figure shows what
the light curve would look like if the source had the same brightness,
despite being much smaller.  In this case, there are quite 
clear (A,B) or relatively clear (C) signatures of an anomaly.
The difference between these two classes is simply the result
of where the gaps fall relative to the strongest part of the anomaly.

The middle panel in each Figure shows a more realistic situation.  The source
is fainter by a factor of 20, similar to a bulge turnoff star.
In this case, there is no recognizable anomaly at all for model C, and
a only a suggestive hint of an anomaly for models A and B.  In any case,
it would not be possible to claim detection of a planet in any of the
three cases.

The bottom panel in each figure shows the whole light curve.  
It is far from clear that the parent microlensing event would even be 
recognized in any of the three cases.

To construct the top panels of these figures, 
we used the original error bars from
Figure~\ref{fig:lc}, and we took the residuals from the zoom
of Figure~\ref{fig:lc} and added these to each of the models shown.
For the lower two panels, we multiplied both the error bars and
residuals by a factor 10.

In creating the Hollywood moniker, \citet{gould97} appears to have had
in mind primarily events in which the caustic is fully enveloped by
the source, in that he emphasized all three characteristics: larger
cross-section, brighter sources, and longer duration. The classic
example of such an event is OGLE-2005-BLG-390 \citep{ob05390}.
However, one can also consider cases, like models A and C
presented here, in which the source is comparable to the size of the
caustic but only partially envelops it, as a second sub-class of
Hollywood events.  We have dubbed these sub-classes as ``Cannae'' and 
``von Schlieffen'' events, respectively. Comparing 
Figures~\ref{fig:artificialA}, \ref{fig:artificialB}, and 
\ref{fig:artificialC} with each other and with Figure~\ref{fig:lc},
one sees that
the latter are still of the ``Hollywood''-type in that they retain the
properties of greater source brightness and longer duration. Thus, the
range of such events is larger than that originally proposed by 
\citet{gould97}. Because the probability of detecting a planet in such an event
scales much less strongly with mass ratio than for a typical source,
Hollywood events can play a crucial role in the detection and
characterization of planets, in particular those of low mass.

Hollywood events do have their drawbacks.  \citet{bennett96} showed
that the signal from Earth/Sun mass-ratio planets will be almost
completely ``washed out'' by giant sources.  On the other hand,
\citet{jung14}, taking account of the higher precision measurements
from giant sources, argued that Earth-mass planets would be
more detectable than for smaller sources (except for separations
$s\sim 1$).  This tension can be illustrated
in the present case by noting from Section~\ref{sec:heuristic} that such
a $q=3\times 10^{-6}$ planet would generate a $\Delta A= 2q/\rho^2 =
0.06$ bump in a fully-enveloped, Cannae event (model B).  The probability for
envelopment would be a factor 10 larger than for a caustic crossing
for a typical, turn-off star, microlensing source.  However, while such a
bump would certainly be detectable in the present data, whether
it could be unambiguously interpreted is less clear.

In any case, the range of $q$ that is accessible to this approach
extends at least a factor 5 below the previous lowest values,
even if it does not reach the Earth/Sun regime.

\begin{figure}
\plotone{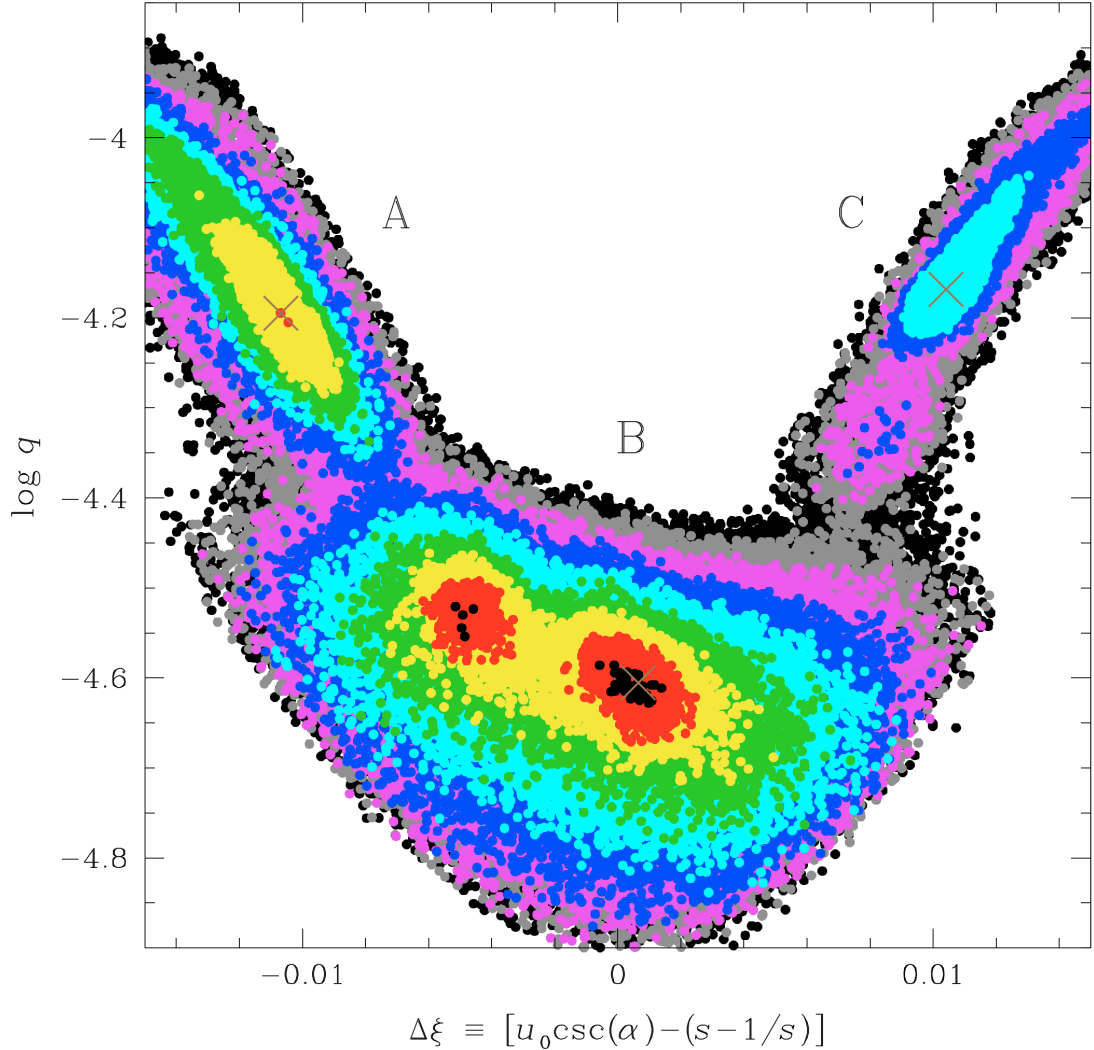}
\caption{Scatter plot of MCMC of 
$\Delta\xi \equiv u_0\csc\alpha - (s - s^{-1})$ vs. $\log q$,
where $\Delta\xi$ is the offset of the center of the source from the
center of the caustic at the moment that the source crosses the binary axis, and
$q$ is the mass ratio.  The plot is derived primarily from a ``hot chain'', to
enable the sampling to cross the $\Delta\chi^2$ barriers between the
three local minima, which we find to be $\Delta\chi^2\sim 35$ between
models A and B, and $\Delta\chi^2\sim 60$ between models B and C.  However,
points from a normal-temperature chain are added to better articulate
the minima.
Color coding is (black, red, yellow, green, cyan, blue, magenta, gray) for
$\Delta\chi^2<(1,4,9,16,25,36,49,64)$.  Values of $\Delta\chi^2>64$ are
again plotted in black.
}
\label{fig:deltaxi}
\end{figure}

\begin{figure}
\plotone{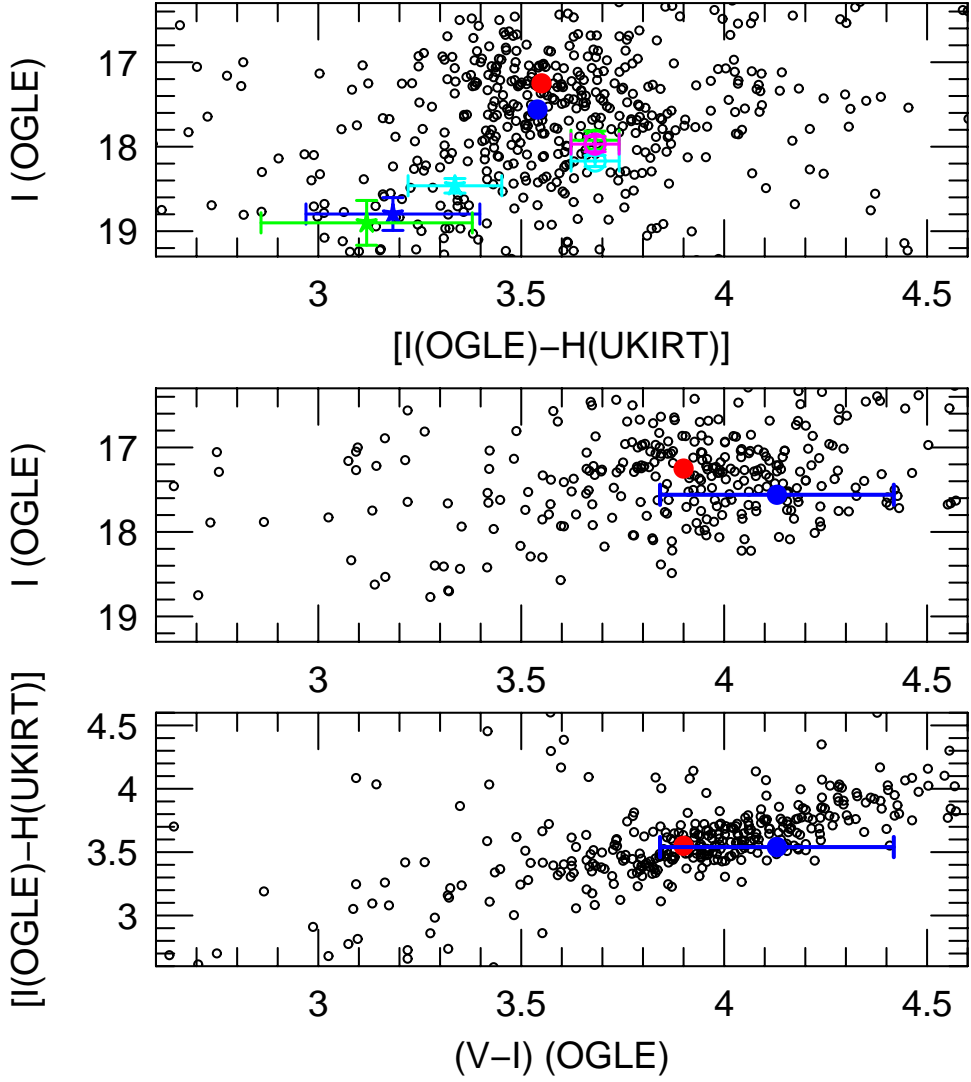}
\caption{CMDs (in $(I-H)$ and $(V-I)$) and $VIH$ color-color diagram
of field stars near OGLE-2017-BLG-0173, together with
the positions of the 
clump centroid (red), the ``baseline object'' (blue), and for each of
the three solutions (A=magenta, B=green, C=cyan), the
source star (open circles),  and the blended 
light (five-pointed stars).  
The ``baseline object'' is barely detected in $V$ band,
which results in a 0.3 mag uncertainty in its color.  
Hence, we use an $[I,(I-H)]$
CMD (top panel) to determine the intrinsic source color.  The offset 
between the source and the clump (in both color and magnitude),
leads to an angular source radius $\theta_*=(5.30,5.43,4.86)\,\muas$, which
is used to estimate the Einstein radius
$\theta_\e=(0.48,0.54.53)\,\mas$ for solutions (A,B,C).  
See Section~\ref{sec:cmd}.
}
\label{fig:cmd}
\end{figure}

\begin{figure}
\plotone{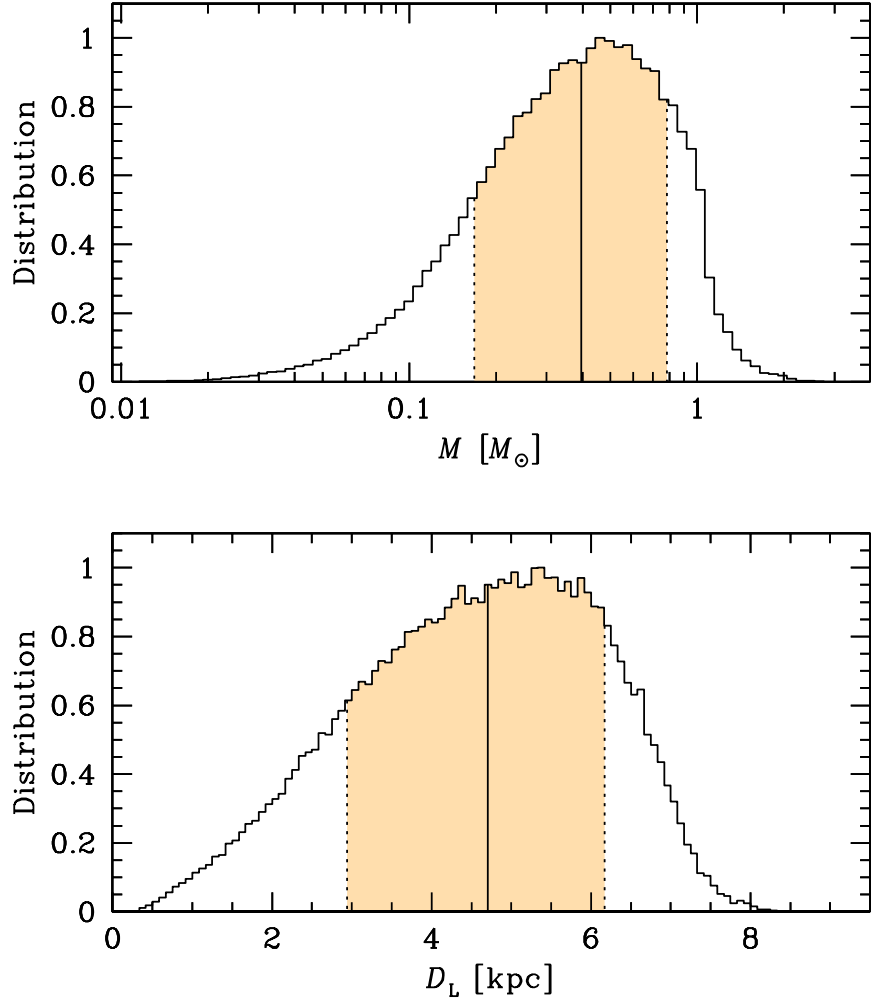}
\caption{Histogram of posterior probabilities of the physical
parameters of the lens system OGLE-2017-BLG-0173L, obtained 
by drawing event parameters from a Galactic model and comparing their
observables $\theta_\e$ and $\mu$ with those derived from 
the microlensing light curve analysis.  This histogram is for model
B, but models A and C are virtually identical.
}
\label{fig:hist}
\end{figure}

\begin{figure}
\plotone{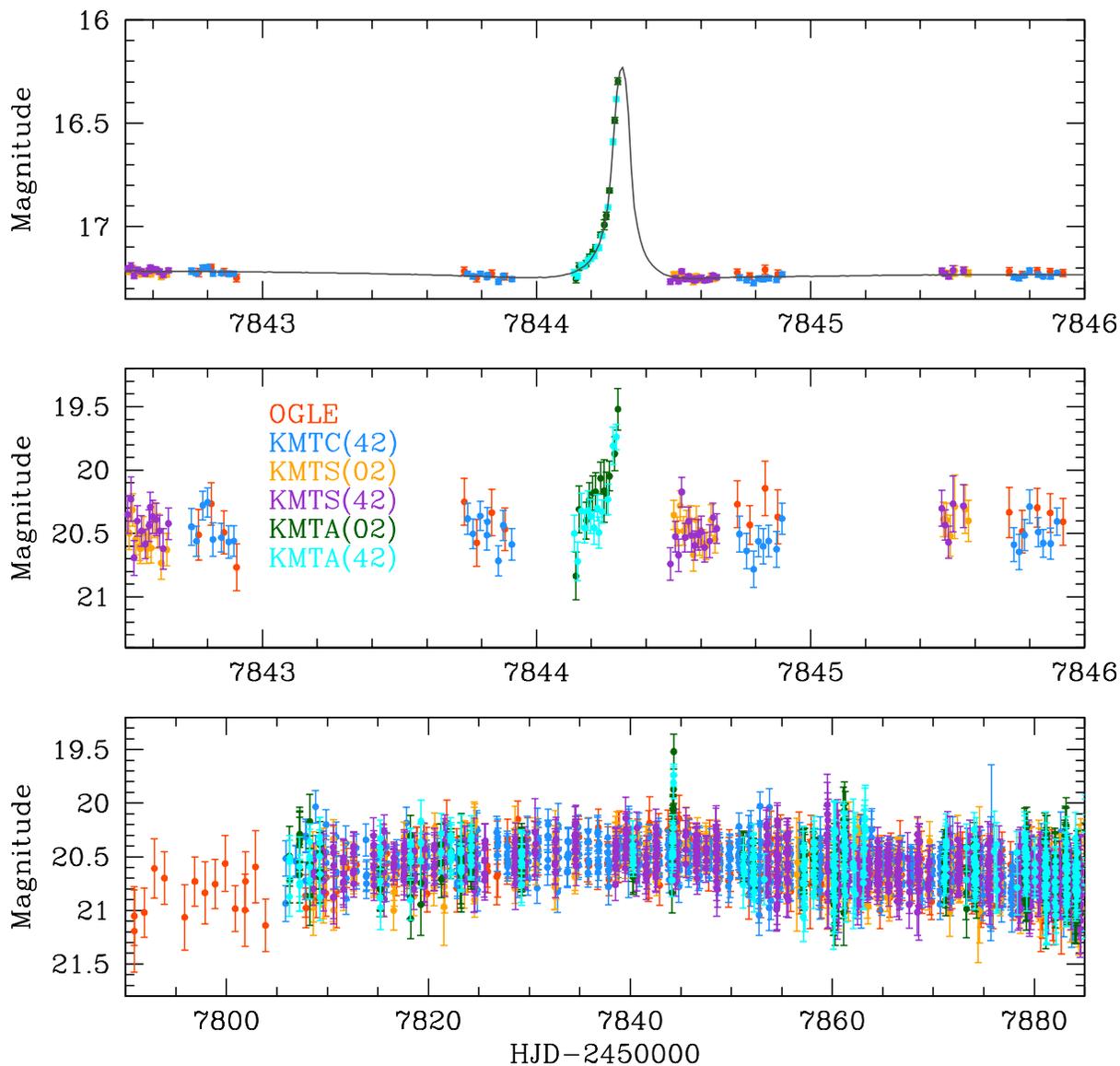}
\caption{Dissection of the virtues of Hollywood through simulated
data, illustrated for model A.  The upper panel shows
how OGLE-2017-BLG-0173 would appear if the source had the same brightness
but were 9.6 times smaller, i.e., similar to typical turnoff star
microlensing sources.  The source passes over the tip of the
cusp, and becomes highly magnified, the rise of which is well-captured
by KMTA data.  In this idealization, the event would be well characterized.
However, the middle panel shows a
more realistic version, in which the source is not only smaller but
also 20 times fainter.  There is only a hint of an anomaly, and this
certainly could not be characterized.  Indeed,
from the bottom panel, it is far from clear that the microlensing event
due to the host star would even be recognized.
}
\label{fig:artificialA}
\end{figure}

\begin{figure}
\plotone{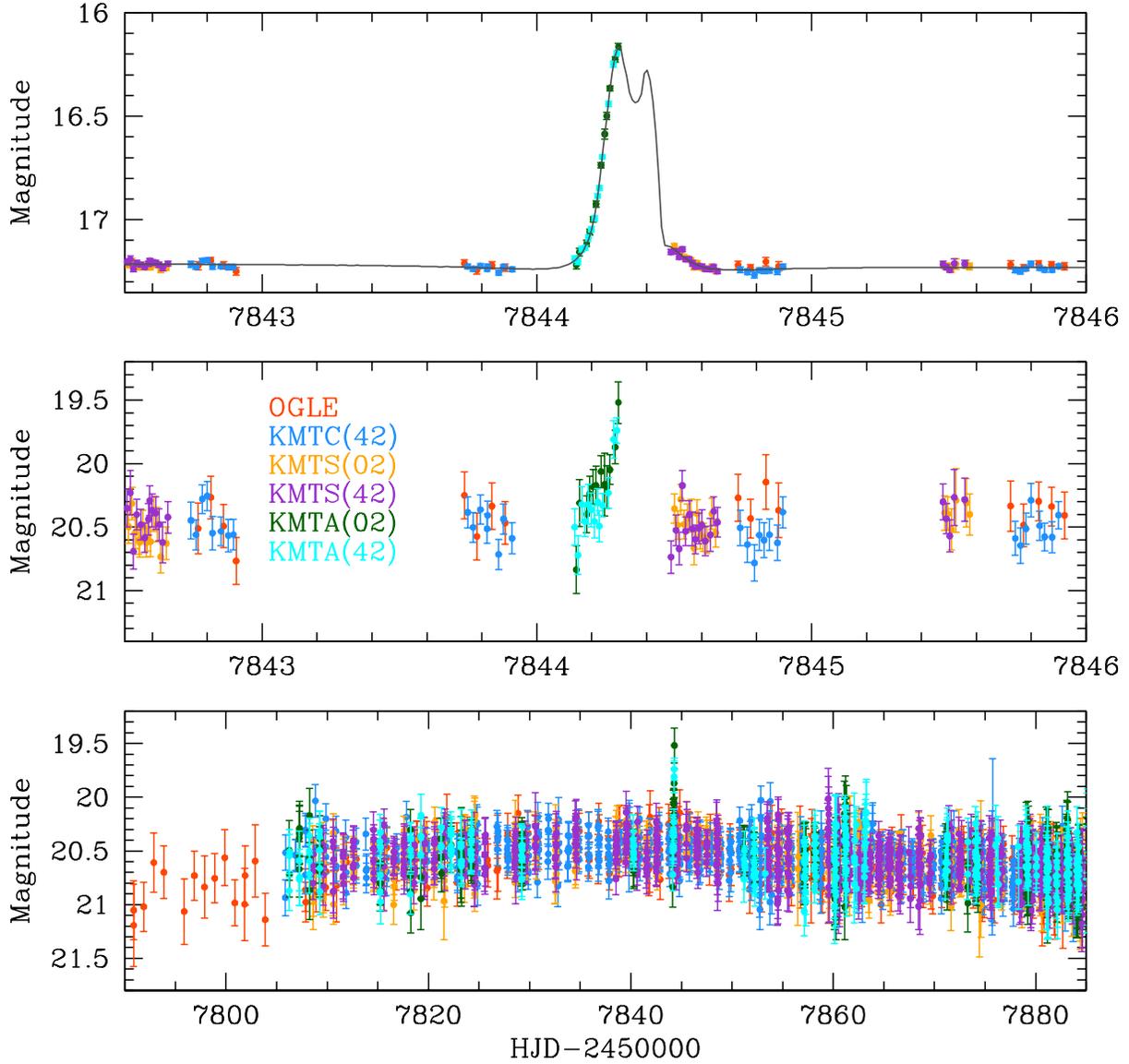}
\caption{Dissection of the virtues of Hollywood through simulated
data, illustrated for model B.  Similarly to model A in 
Figure~\ref{fig:artificialA}, the idealized upper panel would enable
a well characterized planet but the more realistic middle panel (based on
making the source not only smaller but correspondingly fainter) would not.
Again, it is far from clear from the bottom panel that the underlying
event would be recognized as microlensing.
}
\label{fig:artificialB}
\end{figure}

\begin{figure}
\plotone{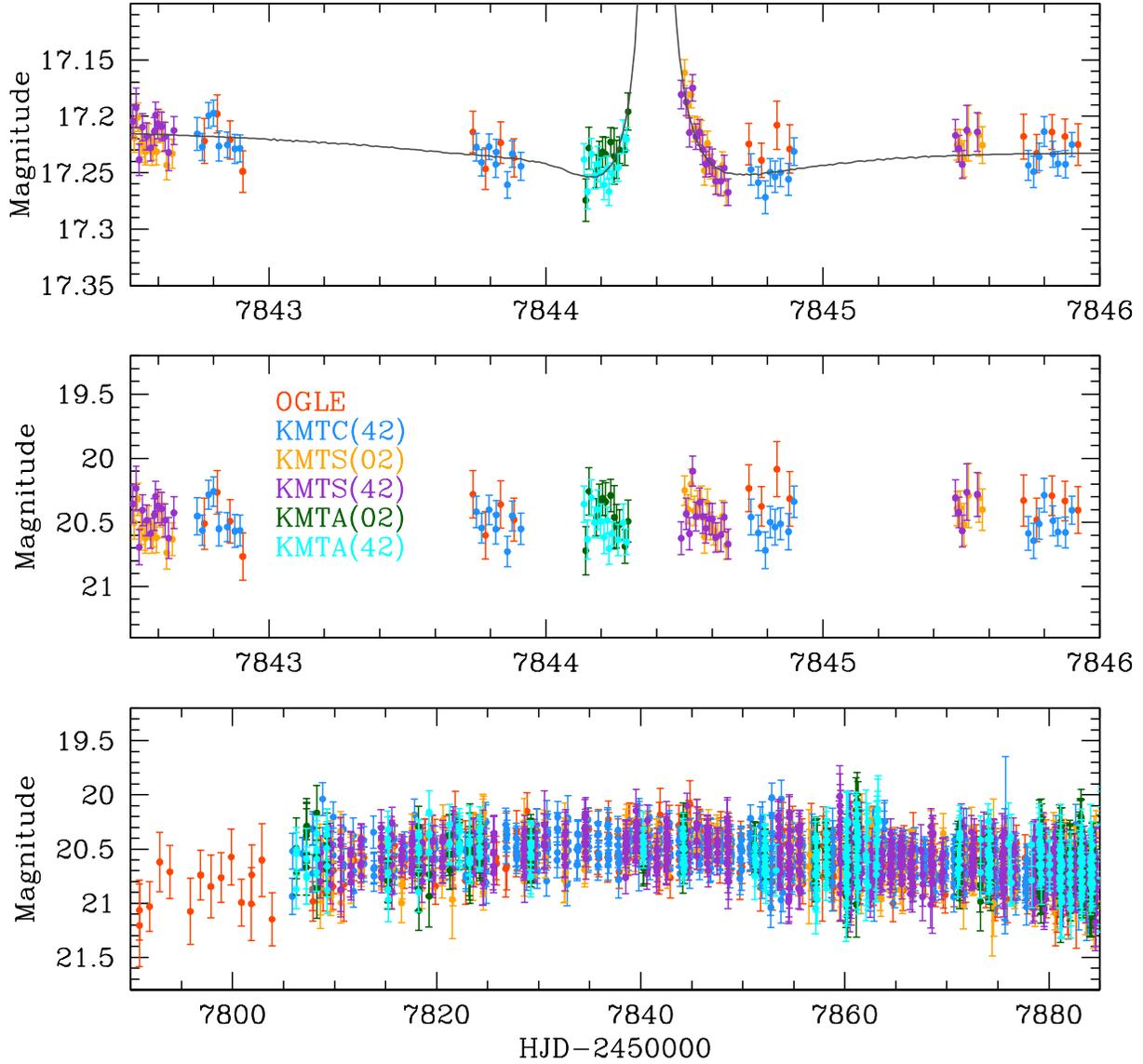}
\caption{Dissection of the virtues of Hollywood through simulated
data, illustrated for model C.  Compared to models 
A (Figure~\ref{fig:artificialA}) and B (Figure~\ref{fig:artificialB}),
the idealized upper panel is substantially less interprepretable due
to the fact that, by chance, almost the entire bump lies in the data
gap.  Hence, in the more realistic middle panel, there is hardly a
hint of the anomaly.
}
\label{fig:artificialC}
\end{figure}

Another potential drawback of Hollywood is the degeneracy discovered in
this paper between Cannae and von Schlieffen solutions.  Although
this degeneracy was shown to be ``accidental'' in the sense that
it was due to data gaps, such gaps are likely to be common.  Furthermore,
this degeneracy is present despite the $\Delta\chi^2\sim 10,000$
detection of the planet.  We therefore investigated the case
of OGLE-2005-BLG-390 \citep{ob05390}, which was the first Hollywood planet.
We find that while the analog of Figure~\ref{fig:deltaxi} shows the
same ``U'' shaped structure, it does not contain multiple minima.
Hence, there is no degeneracy despite the fact that the planet
is detected at only $\Delta\chi^2\sim 500$.  Noting that the source
is much larger than the caustic in the case of OGLE-2005-BLG-390,
we conjecture that this is the decisive difference.  That is, we
suggest that the degeneracy found in OGLE-2017-BLG-0173 will occur
primarily in events for which the source and caustic have similar size.

When the ``Hollywood strategy'' was proposed, it was indeed
a ``strategy'' in the sense that one had to choose to which
targets one should apply limited follow-up resources.  By contrast,
OGLE-2017-BLG-0173Lb was discovered in pure survey mode, in which
no decisions were needed or made about individual targets.
However, the problem of applying limited follow-up telescope resources
does continue to apply to {\it Spitzer} microlensing, and after
the planetary nature of OGLE-2017-BLG-0173 was recognized, the {\it Spitzer}
team \citep{prop2016} revised their selection strategy to give much
greater emphasis to Hollywood events.
More generally, there remains the question of how to apply limited
human resources.  We suggest that searches for smooth bumps, even of
quite low amplitude, in current light curves and archival microlensing
events with giant sources, may yield low-mass outlying planets that
have not previously been recognized.

\acknowledgments 
Work by WZ, YKJ, and AG were supported by AST-1516842 from the US NSF.
WZ, IGS, and AG were supported by JPL grant 1500811.  
Work by C.H. was supported by the grant (2017R1A4A1015178) of
National Research Foundation of Korea.
This research has made use of the KMTNet system operated by the Korea
Astronomy and Space Science Institute (KASI) and the data were obtained at
three host sites of CTIO in Chile, SAAO in South Africa, and SSO in
Australia.
The OGLE project has received funding from the National Science Centre, 
Poland, grant MAESTRO 2014/14/A/ST9/00121 to AU.
Work by YS was supported by an appointment to the NASA Postdoctoral Program at the Jet Propulsion Laboratory,
California Institute of Technology, administered by Universities Space Research Association
through a contract with NASA.
The United Kingdom Infrared Telescope (UKIRT) is supported by NASA and
operated under an agreement among the University of Hawaii, the University
of Arizona, and Lockheed Martin Advanced Technology Center; operations are
enabled through the cooperation of the Joint Astronomy Centre of the Science
and Technology Facilities Council of the U.K.

\begin{deluxetable}{lccc}
\tablecolumns{4} 
\tablewidth{0pc}
\tablecaption{\textsc{Best-fit Solution}} 
\tablehead{ 
   \colhead{Parameters } & 
   \colhead{model A} &
   \colhead{model B} &
   \colhead{model C}
} 
\startdata
$\chi^2/\rm{dof}$             & 7445.54/7443       & 7442.07/7443       & 7458.08/7443       \\
$t_0$ $(\rm{HJD}^{\prime})$   & 7838.031$\pm$0.059 & 7837.946$\pm$0.059 & 7838.011$\pm$0.059 \\
$u_0$                         &    0.844$\pm$0.035 &    0.867$\pm$0.043 &    0.768$\pm$0.028 \\
$t_{\rm E}$ $(\rm{days})$     &   30.818$\pm$0.898 &   30.460$\pm$1.040 &   32.930$\pm$0.846 \\
$s$                           &    1.532$\pm$0.025 &    1.540$\pm$0.031 &    1.465$\pm$0.019 \\
$q$ $(10^{-5})$               &    6.386$\pm$1.001 &    2.479$\pm$0.242 &    6.788$\pm$0.729 \\
$\alpha$ $(\rm{rad})$         &    1.334$\pm$0.004 &    1.332$\pm$0.004 &    1.324$\pm$0.003 \\
$\rho$ $(10^{-3})$            &   10.969$\pm$0.469 &   10.024$\pm$0.512 &    9.150$\pm$0.331 \\
$F_s$                          &    1.144$\pm$0.093 &    1.198$\pm$0.119 &    0.957$\pm$0.064 \\
$F_b$                          &    0.391$\pm$0.093 &    0.337$\pm$0.119 &    0.578$\pm$0.064 \\
\enddata 
\label{tab:ulens}
\end{deluxetable}

\begin{deluxetable}{lccc}
\tablecolumns{4} 
\tablewidth{0pc}
\tablecaption{\textsc{Physical properties}} 
\tablehead{
   \colhead{Quantity} &
   \colhead{model A} & 
   \colhead{model B} &
   \colhead{model C}
}
\startdata
$M_{\rm host}$ $[M_\sun]$      & $0.357_{-0.208}^{+0.360}$  & $0.396_{-0.227}^{+0.390}$ & $0.396_{-0.229}^{+0.386}$  \\
$M_{\rm planet}$ $[M_\earth]$  & $7.581_{-4.911}^{+10.045}$ & $3.269_{-2.015}^{+3.849}$ & $8.950_{-5.585}^{+10.626}$ \\
$D_{\rm L}$ [kpc]              & $5.015_{-1.848}^{+1.463}$  & $4.705_{-1.763}^{+1.468}$ & $4.800_{-1.793}^{+1.463}$  \\
$a_\bot$ [AU]                  & $3.688_{-1.540}^{+1.456}$  & $3.913_{-1.649}^{+1.615}$ & $3.727_{-1.553}^{+1.478}$  \\
\enddata
\label{tab:phys}
\end{deluxetable}

\end{document}